\documentclass[aps,pre,onecolumn,10pt]{revtex4-1}
\usepackage{amsmath}
\usepackage{graphicx}
\usepackage{color}
\usepackage{hyperref}
\usepackage{psfrag}
\usepackage{stmaryrd}
\usepackage{amssymb}
\usepackage{wasysym}
\usepackage{multirow}

\usepackage{geometry}
\newcommand{\cor}[1]{\textcolor{black}{#1}}

\def \pd{\partial}
\def \be{\begin{equation}}
\def \ee{\end{equation}}
\def \ba{\begin{eqnarray}}
\def \ea{\end{eqnarray}}

\begin{document}

\title{Mean mass transport in an orbitally shaken cylindrical container}
\author{Julien Bouvard$^1$}
\author{Wietze Herreman$^2$}
\author{Fr\'ed\'eric Moisy$^1$}
\affiliation{$^1$Laboratoire FAST, Univ. Paris-Sud, CNRS, Universit\'e Paris-Saclay, 91405 Orsay, France.}
\affiliation{$^2$Laboratoire LIMSI, CNRS, Univ. Paris-Sud, Universit\'e Paris-Saclay, 91405 Orsay, France.}

\date{\today}

\begin{abstract}

\cor{A cylindrical container partially filled with a liquid} in orbital shaking motion, i.e. in circular translation with fixed orientation with respect to an inertial frame of reference, generates, along with a rotating sloshing wave, a mean flow rotating in the same direction as the wave. Here we investigate experimentally the structure and the scaling \cor{of the wave flow and the Lagrangian mean flow} in the weakly nonlinear regime, for small forcing amplitude and for forcing frequency far from the resonance, using conventional and stroboscopic particle image velocimetry. \cor{The Lagrangian mean flow is composed of a strong global rotation near the center and a non trivial pattern of poloidal recirculation vortices of weaker amplitude, mostly active near the contact line. The global rotation near the center is robust with respect to changes in viscosity and forcing frequency, and its amplitude compares well with the predicted Stokes drift for an inviscid rotating sloshing wave. On the other hand, the spatial structure of the poloidal vortices show strong variation with viscosity and forcing frequency, suggesting that it results from the streaming flow driven by the complex oscillatory boundary layers near the contact line.
}

\end{abstract}

\maketitle

\section{Introduction} \label{sec:intro}

It is common knowledge that prescribing an orbital motion to a glass of wine generates a rotating gravity wave  that comes along with a  swirling mean flow~\cite{Hutton1964}. This mean flow rotates in the direction of the wave and recirculates poloidaly (radially and vertically), thus permanently pushing new fluid to the surface where it aerates and releases the wine's aromas~\cite{ReclariPhD,Reclari2014}. Precisely the same kind of orbital shaking \cor{is used in} bioreactors for the cultivation of biological cells~\cite{Wurm2013}. There, the presence of the mean flow prevents sedimentation and ensures efficient gas exchange, avoiding the damagingly high shear rates that immersed stirrers would cause.

Because of its importance in engineering applications, experimental and numerical efforts have been made to optimize the mixing efficiency and power consumption of orbital shakers~\cite{Gardner1992,Buchs2000,Peter2006}.  Most of the studies focus on the strongly nonlinear regime (forcing frequency close to the fundamental resonance frequency of the container), and in complex container geometries such as Erlenmeyer flasks with baffles.  Far from the resonance, the forced rotating wave is well described by inviscid linear potential theory, and simply corresponds to a superposition of \cor{two normal linear sloshing modes with $\pi/2$ phase shift}~\cite{Hutton1964,Ibrahim2005,Faltinsen2014,Reclari2014}. Near resonance, the wave becomes large and displays complex nonlinear phenomena such as wave breaking and hysteresis~\cite{Reclari2014}. Note that a rotating wave can also be triggered near resonance in a linearly shaken container through a symmetry breaking mechanism~\cite{Dodge2000,Funakoshi1988,Royon2007}. Experimental advances using particle image velocimetry (PIV) measurements in the frame of the container have recently opened the way to quantitative mixing diagnostics in this system (turbulent intensity, local energy dissipation rate)~\cite{Weheliye2013, Ducci2014}.  

In recent years, numerical and experimental efforts have been devoted to understand the orbital sloshing flow in a simple cylindrical container on a more fundamental level~\cite{Kim2009,Reclari2014}.   In spite of these efforts, no general picture is available yet for the mechanism that generates the mean flow induced by orbital shaking and its dependence with the flow parameters (aspect ratio of the container, fluid viscosity, forcing amplitude).

\cor{Mean flows driven by nonlinear interaction of} surface waves is a classical problem of fluid mechanics that received much attention mainly because of its oceanographical interest --- see for example Refs.~\cite{Longuet1953,Batchelor1967,Craik76,Craik1982} for historical work on long propagative gravity waves, or Refs.~\cite{Higuera2005,Perinet2017} for more recent studies on parametrically excited capillary-gravity waves. Here, we focus on the weakly nonlinear limit, where the amplitude of the waves, harmonics and mean flows can be ranked in decreasing orders of magnitude. \cor{When discussing mean flows, it is important to distinguish the Eulerian mean flow from the Lagrangian mean flow. The Eulerian mean flow corresponds to the time-averaged value of the velocity field, and is commonly called the streaming $\overline{\bf u}_{str}$ (or {\it steady streaming} when $\overline{\bf u}_{str}$ is stationary). The Lagrangian mean flow $\overline{\bf u}= \overline{\bf u}_{str} + \overline{\bf u}_{Sto}$ corresponds to the flow that induces the mean transport of mass. It has two contributions, the streaming $\overline{\bf u}_{str}$ and an additional purely kinematic contribution, the Stokes drift $\overline{\bf u}_{Sto}$. In this article, the term mean flow denotes mean transport of mass $\overline{\bf u}$, so it is important to address both $\overline{\bf u}_{str}$ and $\overline{\bf u}_{Sto}$.}

In incompressible fluids, streaming is essentially the reaction of the flow to the time-averaged non-linear advection \cor{$\overline{{\bf u} \cdot \nabla {\bf u}}$, where ${\bf u}$} denotes the linear wave flow~\cite{Batchelor1967,Riley2001}.  In the case of inviscid potential gravity waves, this basic principle becomes rather subtle, because such waves do not carry vorticity and as such cannot generate a \cor{streaming} flow in the bulk. Only in the viscous boundary layers over the container walls and below the liquid surface a non-zero forcing of a streaming flow can take place, and theoretical modeling \cor{of nonlinear interactions therein} becomes highly non-trivial in three-dimensional contexts~\cite{Nicolas2003}. An estimate for the steady streaming magnitude can be obtained as follows. The oscillatory (Stokes) boundary layers have a typical thickness $\delta = (\nu / \Omega)^{1/2}$, with $\Omega$ the wave frequency and $\nu$ the kinematic viscosity. When the amplitude (or phase) of the wave varies along that boundary, mass conservation requires that there must exist a small oscillating flow normal to the boundary. This flow, called boundary layer pumping, is of order $u \delta/L$, where $L$ is the length-scale over which the wave magnitude $u$ varies. Since the tangential and normal velocities in the boundary layer are not out of phase in general, their average product is non-zero, which induces an effective Reynolds stress of order $u^2 \delta/L$ that forces a non-zero mean flow in the bulk.  The structure of this streaming flow depends on the details of the boundary conditions, but its magnitude \cor{$\overline{u}_{str}$} can be inferred, at least in the weak steady streaming limit (i.e., for small streaming Reynolds number \cor{$\overline{u}_{str} L / \nu \ll 1$}), from a balance between the Reynolds stress and the viscous stress \cor{$\nu \overline{u}_{str} / \delta$} at the boundaries, yielding
\cor{$ \overline{u}_{str}  \sim {u^2}/{\Omega L}.$} Interestingly, this amplitude is independent of viscosity, although viscosity is a necessary ingredient for the formation of the boundary layers from which the streaming originates. We recall that this estimate only applies in the weak streaming limit, for which the wave properties remain unaffected by the weak mean flow. In the strong streaming regime, the streaming flow can alter the waves, and inertial effects must be considered to solve for the mean flow dynamics~\cite{Craik76}.

\cor{The second contribution to the mean mass transport, the Stokes drift, has a purely kinematic origin that can be understood as follows \cite{Stokes1847}.} Fluid particles in oscillatory flows move along \cor{nearly circular} trajectories and experience, along these paths, slight variations in the wave magnitude. This causes them to displace slightly more in one direction and leads to a small mean drift after each wave period. A noticeable property of the Stokes drift is that it allows mean mass transport even in the absence of \cor{an Eulerian mean flow (when $\overline{{\bf u}}_{str} = {\bf 0}$)}. An estimate for the Stokes drift magnitude \cor{$\overline{u}_{Sto}$} can be simply obtained by multiplying the gyration radius of the particles $u / \Omega$ with the spatial gradient of the wave amplitude $u / L$,  \cor{which brings us the estimate $\overline{u}_{Sto}  \sim u^2/ \Omega L$. The amplitude of the Stokes drift therefore has the same order of magnitude as that of the streaming.}

\cor{Summarizing, we can expect that the magnitude of the Lagrangian mean flow $\overline{\bf u}$ that controls the mean transport of mass behaves as
\be
\overline{u} \sim \frac{u^2}{\Omega L}. \label{eq:scalsstr}
\ee
Because of the identical scaling of the two contributions $\overline{\bf u}_{str}$ and $\overline{\bf u}_{Sto}$, it is difficult in general to anticipate which of streaming or Strokes drift contribute more to this Lagrangian mean flow~\cite{Larrieu2009}.}

\cor{Predicting the structure of the Lagrangian mean flow in the orbital sloshing configuration is a difficult task, not only because of the non-trivial viscous boundary layers that generate the streaming, but also because of the complex dynamics near the contact line.} The above scaling argument (\ref{eq:scalsstr}) can however be readily applied to predict its amplitude, at least in the weakly nonlinear regime. In a cylinder of radius $R$, orbitally shaken along a path with circular radius $A$, and for forcing frequency $\Omega$ small compared to the natural frequency $\omega_1$ of the first sloshing mode of the cylinder, the non-dimensional wave amplitude can be estimated using potential theory as~\cite{Ibrahim2005,Faltinsen2014,Reclari2014},
\begin{equation}
\frac{u}{\Omega R} \sim \frac{\epsilon}{(\omega_{1}/\Omega)^2 - 1},
\label{eq:scaluw}
\end{equation}
where $\epsilon=A/R \ll 1$ is the non-dimensional forcing amplitude. According to Eq.~(\ref{eq:scalsstr}), this wave is therefore expected to generate a \cor{Lagrangian mean flow $\overline{{\bf u}} = \overline{{\bf u}}_{str}+\overline{{\bf u}}_{Sto}$} of magnitude
\begin{equation}
\frac{\overline{u}}{\Omega R}  \sim \frac{\epsilon^2}{[(\omega_{1}/\Omega)^2 - 1]^2}.
\label{eq:scalub}
\end{equation} 
This indicates that the mean flow magnitude increases very rapidly with $\Omega$, at least as $\overline{u} \sim \Omega^5$ far from the resonance, and that it does not depend on viscosity. This scaling is compatible with the calculation of of Hutton~\cite{Hutton1964} \cor{who considered only the Stokes drift associated to the potential flow solution}. However, the Stokes drift solution of Hutton corresponds to a purely azimuthal mean flow, lacking the poloidal recirculations observed experimentally, so it cannot provide a complete description of the complex mean flow patterns induced by orbital sloshing.

In this paper, we present a systematic series of experiments to determine the structure and the scaling of the wave flow and the \cor{Lagrangian} mean flow in the weakly nonlinear regime. The mean flow is determined using stroboscopic PIV, i.e. PIV synchronized with the forcing, which naturally filters out the wave motion. We carefully discuss some key aspects of stroboscopic PIV: \cor{being a particle based method, it measures the Lagrangian mean flow and therefore cannot discriminate the Stokes drift from the streaming contribution. Moreover, mean flows measured by stroboscopic PIV are} affected by a systematic bias due to the arbitrary phase of the acquisition; \cor{we show how this bias can be removed by making use of the axisymmetry of the mean flow}. By varying the forcing amplitude $\epsilon$ and frequency $\Omega/\omega_1$, we show that the amplitude of the global rotation near the center is well described by Eq.~(\ref{eq:scalub}) with weak influence of viscosity, and that its structure is compatible with a dominant Stokes drift contribution. On the other hand, we find that the spatial structure of the poloidal vortices mostly active near the contact line show strong variations with viscosity and forcing frequency. \cor{This suggests a strong streaming response to the oscillating boundary layer near the contact line, a feature that is neglected in all the available theories on streaming induced by surface waves.}

\section{Experimental set-up and procedure}

\begin{figure}
    \centerline{\includegraphics[width=14cm]{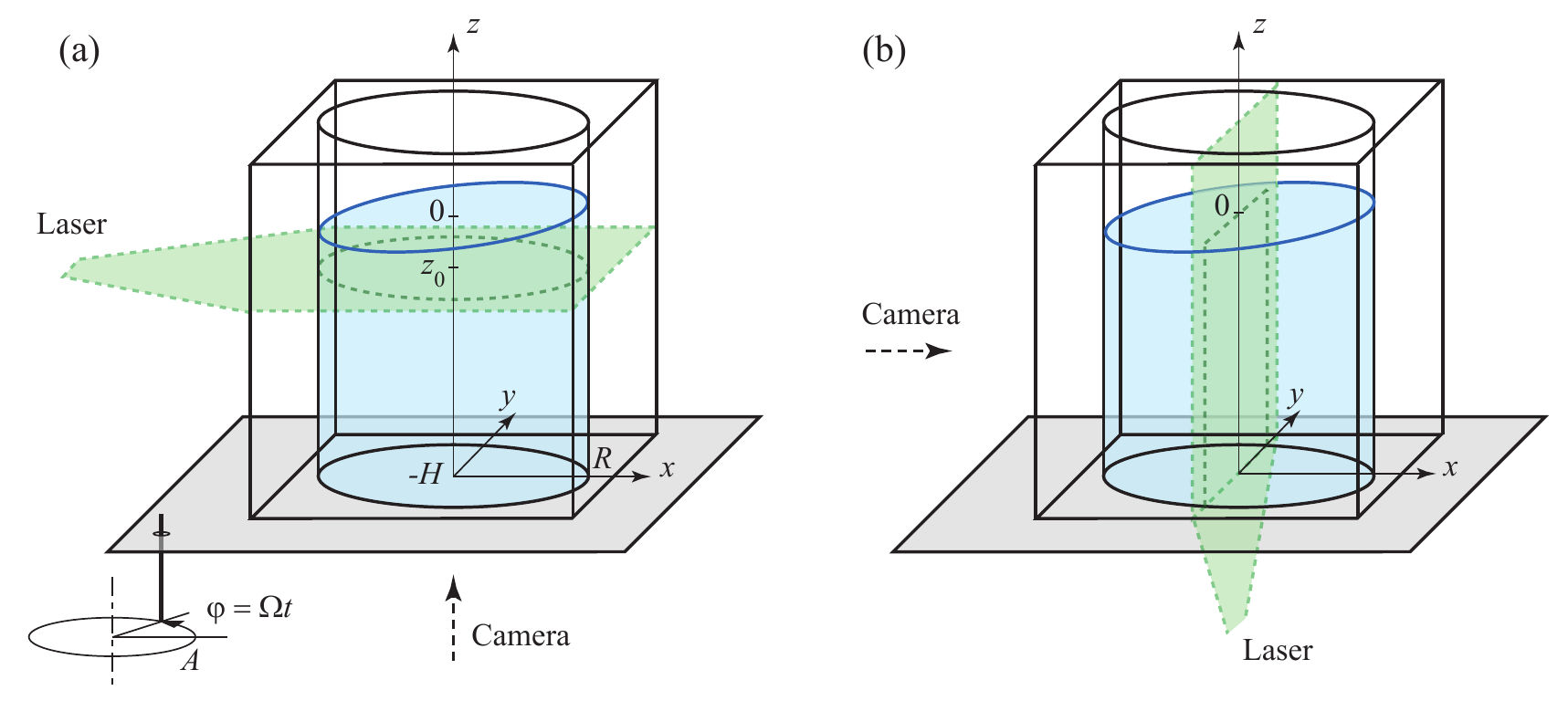}}
    \caption{Experimental setup. The cylinder of radius $R=51.2$~mm is filled up to a height $H=111$~mm by silicon oil of kinematic viscosity $\nu$.  The entire system is oscillated at a constant frequency $\Omega$ along a circular trajectory of radius $A$, maintaining a fixed orientation with respect to an inertial frame of reference. The particle image velocimetry measurements are performed using a laser sheet (dashed green) and a camera fixed in the laboratory frame. (a) Horizontal measurements in the plane $z_0/R = -0.23$, for the wave flow and the mean flow; (b) Vertical measurements, for the mean flow only.}
		\label{fig:setup}
\end{figure}

The experimental set-up is sketched in Fig.~\ref{fig:setup}. A glass cylinder of inner radius $R = 51.2$~mm is filled at height $H = 111$~mm with silicon oil of kinematic viscosity \cor{either} $\nu = 50$ or 500~mm$^2$~s$^{-1}$ and surface tension $\gamma =  21 \times 10^{-3}$~N~m$^{-1}$. \cor{The cylinder is located on a transparent plate attached to an eccentric motor that shakes the entire system at angular frequency $\Omega$ along a circle of radius $A$. The motion of the plate is constrained by two linear guide rails in the $x$ and $y$ directions, ensuring a pure circular translation with no rotation component.} We choose the origin of time such that an arbitrary point of the container follows the path ${\bf r}_c (t) = {\bf r}_c (0) +  A (  \cos  \Omega t\, {\bf e}_x  + \sin  \Omega t \, {\bf e}_y )$. In the co-moving frame attached to the cylinder, this induces an effective gravitational acceleration ${\bf g}'(t) = -g {\bf e}_z + A \Omega^2 (  \cos  \Omega t\, {\bf e}_x  + \sin  \Omega t \, {\bf e}_y )$.

The shaking amplitude $A$ is varied between 0.3 and 10~mm with precision $0.1$~mm and the shaking frequency $\Omega$ can be set between 0 and 270~rpm with precision 0.1~rpm.  However, measurements for $\Omega<90$~rpm were found to be hindered by weak fluid motions, of the order of $0.2$~mm~s$^{-1}$, due to residual thermal convection in the cylinder, so measurements are restricted to $\Omega>90$~rpm in the following. \cor{The surface elevation of the rotating wave is of the order of 1~mm for $\Omega = 90$~rpm, and reaches 15~mm near the resonance.  The capillary length (and hence the typical size of the meniscus) is $\lambda_c = \sqrt{\gamma/\rho g} \simeq 1.5$~mm. The good wetting of the oil on the glass wall ensures that the contact line follows the wave motion with minimum pinning effect.}

The system is characterized by five non-dimensional numbers,
\ba
h=  \frac{H}{R}= 2.17, \qquad  \epsilon = \frac{A}{R} \in [0.006, 0.20], \qquad  Bo = \frac{\rho g R^2}{\gamma } = 1100 \nonumber \\
 Re = \frac{\Omega R^2}{\nu} \in [50,1500], \qquad    Fr = \frac{A \Omega^2}{g}  \in [0.0027, 0.8],
\label{eq:ndn}
\ea
\cor{corresponding to the aspect ratio of the cylinder $h$, the normalized forcing amplitude $\epsilon$, the Bond number $Bo$, the Reynolds number $Re$ and the Froude number $Fr$. The Reynolds number defined here compares the cylinder radius $R$ to the boundary layer thickness $\delta = (\nu / \Omega)^{1/2}$. Its large value indicates that the wave motion can be considered as essentially inviscid. Note that the Reynolds number based on the expected mean flow amplitude (\ref{eq:scalub}), $Re_s \sim \epsilon^2 Re$, is small for much of our experiments, but it can exceed O(1) for the lower fluid viscosity as the resonance is approached, from which we can anticipate that the mean flow is not necessarily stable (see Sec.~\ref{sec:dmfp}). Finally, the large value of the Bond number, which compares the cylinder radius to the capillary length $\lambda_c$, indicates that the capillary effects can be neglected in the dispersion relation; this however does not imply that the complex dynamics near the contact line can be neglected in the mean flow generation.}

To perform PIV measurements, we seed the fluid with silver-coated neutrally buoyant particles, {and illuminate it} by a pulsed laser sheet, either vertical or horizontal (see Fig.~\ref{fig:setup}). To minimize refraction through the curved wall, the cylinder is immersed in a cubic container, filled with the same silicon oil. For horizontal measurements (Fig.~\ref{fig:setup}a), the flow is imaged from below, and the laser sheet is located at a height $z_0 = -12$~mm below the surface at rest \cor{to avoid intersection of the laser sheet with the tilted free surface}. Images are mirrored in the following to appear in the right coordinate system $(x,y)$. For vertical measurements (Fig.~\ref{fig:setup}b), the laser sheet is emitted from below, and the flow is imaged from the side, in the plane ($y,z)$, {at times $\Omega t_n  = \pi/2 + 2 n \pi$} for which the cylinder axis crosses the laser sheet. 

Velocity fields are computed from PIV using two different schemes for the wave flow and for the mean flow:

(i) The wave flow is measured using conventional PIV, i.e. with a time delay between images {that is} small compared to the forcing period. This method is essentially insensitive to the mean flow, and provides to a good approximation the Eulerian wave flow.

(ii) The mean flow is measured using stroboscopic PIV, i.e. with image acquisition synchronized with the forcing, in order to filter out the wave motion. This method, similar to that used recently by Perinet {\it et al.}~\cite{Perinet2017} in Faraday wave experiments, essentially measures the total Lagrangian mean flow (see Sec.~\ref{sec:mf}).

Since the PIV setup (laser and camera) is in the laboratory frame, measurements of the vertical structure of the wave flow cannot be performed,  so only measurements in a horizontal plane are achieved. On the other hand, for the mean flow, measurements can be performed both in the horizontal plane (mean rotation) and in the vertical plane (mean poloidal recirculation), because the synchronization of image acquisition with the cylinder motion naturally cancels the contribution due to the cylinder velocity.

Care was taken to ensure the damping of transients before PIV acquisition. Preliminary experiments have shown that the convergence of the wave flow {to a stationary regime} is achieved very rapidly, after less than 10 forcing periods for $\nu = 50$~mm$^2$~s$^{-1}$. On the other hand, the convergence of the mean flow is achieved on a much slower time scale, after typically 1000 forcing periods (10 minutes). In the following, we wait at least 10 minutes between each measurement.

\section{Rotating wave flow}
\label{sec:wf}

\subsection{Inviscid potential solution}
\label{sec:pot}

We briefly recall here the main results of the potential theory, obtained by summing two linear sloshing modes with $\pi/2$ phase shift~\cite{Ibrahim2005,Reclari2014}. \cor{Noting $z=0$ at the fluid surface and $z=-H$ at the bottom, the velocity potential in the reference frame of the cylinder reads}
\be
\frac{\phi(r,\theta,z,t)}{\Omega R^2} =  2 \epsilon \sin (\theta - \Omega t) \sum_{n=1}^\infty \alpha_n J_1 (k_{n} r/R)  \cosh [k_{n}(z+H)/R],
\ee
with
\be
\alpha_n = \frac{1}{(k_n^2-1)[(\omega_n / \Omega)^2-1] J_1(k_n) \cosh (k_n H/R)}.
\ee
In the spatial structure of this potential, we recognize the potential of free gravity waves, with azimuthal wavenumber $m=1$ and finer structures in the radial direction as $n$ increases. The numbers $k_n$ are the $n$th zeros of the derivative of $J_1$, the Bessel's function of the first kind and first order ($k_{1} \simeq 1.841$, $k_2 \simeq 5.331$...).  The natural frequencies of these gravity waves are given by
\be
\omega_{n}^2 = \frac{g k_n}{R} \tanh \left( \frac{k_{n} H}{R} \right).
\label{eq:wn}
\ee
The velocity field in the reference frame of the cylinder, ${\bf u} = \nabla \phi$, is
\begin{eqnarray}
\frac{u_r}{\Omega R}      &=  & 2 \epsilon \sin(\theta - \Omega t) \sum_{n=1}^\infty \alpha_n k_n J'_1 (k_n r/R) \cosh [k_n(z+H)/R], \nonumber \\
\frac{u_\theta}{\Omega R} &= &  2 \epsilon \cos(\theta - \Omega t) \sum_{n=1}^\infty \alpha_n k_n \frac{J_1 (k_n r/R)}{k_n r/R} \cosh [k_n(z+H)/R], \nonumber \\
\frac{u_z}{\Omega R}      &= &  2 \epsilon \sin(\theta - \Omega t) \sum_{n=1}^\infty \alpha_n k_n J_1 (k_n r/R) \sinh [k_n(z+H)/R]. 
\label{eq:uth}
\end{eqnarray}
Obviously, the linear potential theory only holds for $\Omega$ far from the natural frequencies $\omega_n$,  since otherwise $\alpha_n \rightarrow \infty$ and viscous or nonlinear effects must be considered to regularize the theory. For low forcing frequencies $\Omega \ll \omega_1$, far enough under the first resonance, the wave is dominated by the first mode $n=1$. We see that in this regime the wave amplitude scales as Eq.~(\ref{eq:scaluw}), as discussed in the Introduction. In the following we will use $(\epsilon, \Omega/\omega_1)$ as control parameters; the normalized frequency $\Omega / \omega_1$ is trivially related to $Fr$, $h$ and $\epsilon$, but is of more practical use. In our set-up, the frequency of the first resonance is $\omega_{1} = 180$~rpm and we can cover the interval $\Omega / \omega_1 \in [0.5, 1.5]$.

Viscous effects are absent in this potential model, but we expect that, in the limit of large $Re$ and far enough from resonance, the inviscid linear solution (\ref{eq:uth}) provides a reasonable description of the wave flow far from the boundaries. Near the wall and under the free surface, boundary layers of thickness $\delta = \sqrt{\nu / \Omega}$ develop in order to meet no-slip and free-surface boundary conditions. We expect that viscous damping therein reduces the wave amplitude and introduces a phase shift with respect to the forcing, as observed in linear sloshing problem~\cite{Bauer1999}.

\subsection{Experimental measurements}

\begin{figure}
    \centerline{
		\includegraphics[width=14cm]{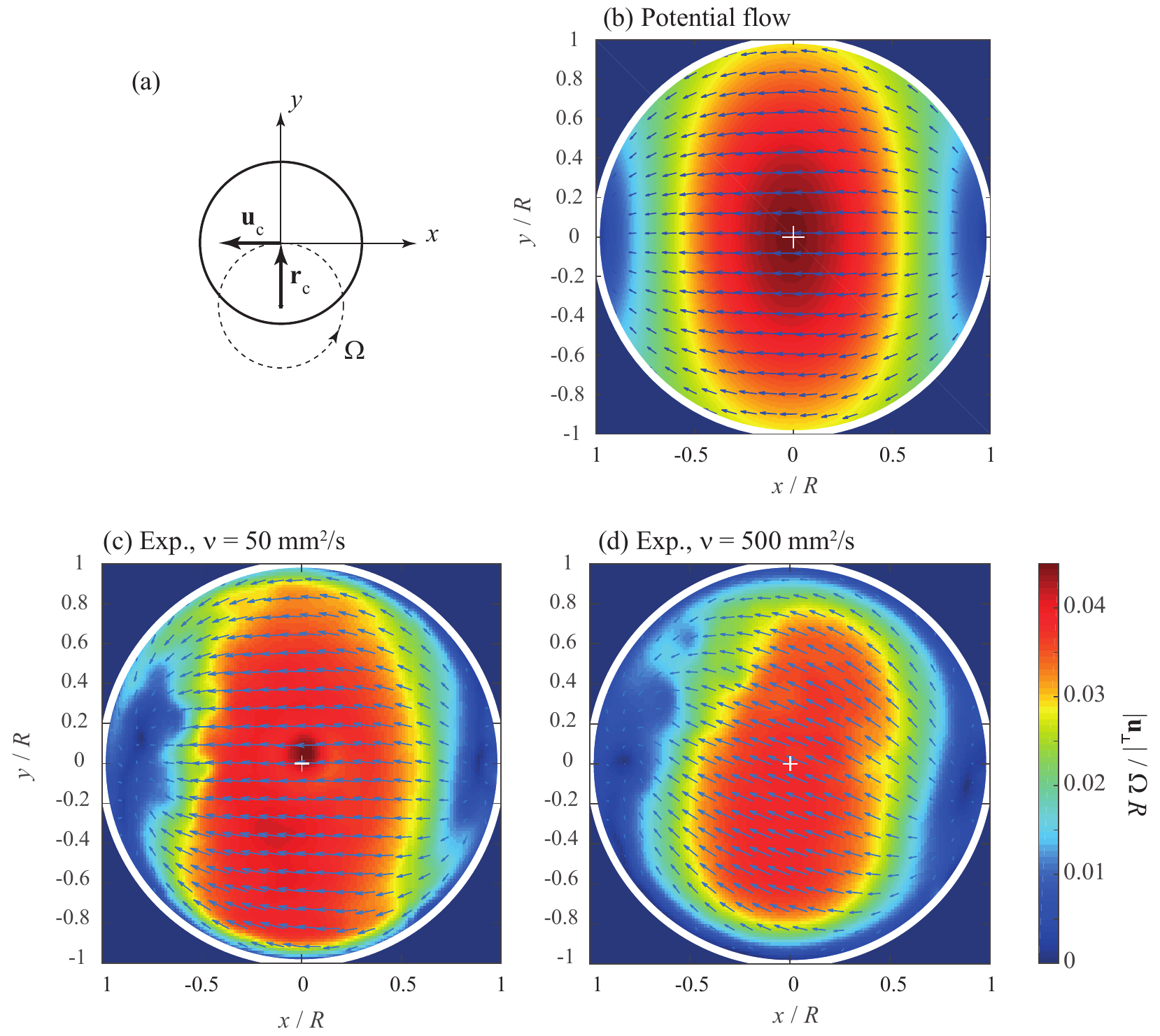}
	}
    \caption{(a) Orbital motion of  the cylinder; the velocity fields in (b)-(d) are taken at phase $\varphi_0 = \pi/2$, for which the cylinder velocity ${\bf u}_c$ is along $-{\bf e}_x$. (b) Potential flow solution. (c,d) Experimental wave field in the frame of the cylinder, measured at a distance $z_0/R=-0.23$ below the surface, for forcing frequency $\Omega/\omega_{1} = 0.67$, forcing amplitude $\epsilon = A/R = 0.057$, and fluid viscosity (c) $\nu = 50$, (d) $\nu = 500$~mm$^2$~s$^{-1}$.  The colormap represents the norm of the horizontal velocity.}
		\label{fig:wave_PIV}
\end{figure}

\cor{To measure the wave component of the flow, we use conventional particle image velocimetry: at each period $n$ of the forcing, two images separated by a small time lag $\delta t$ are acquired, at times $\Omega t_n^{\pm} = \pi/2 + 2\pi n \pm \Omega \delta t/2$. The time lag $\delta t = t_{n}^+ - t_n^-$ is chosen such that the typical particle displacement is a fraction of the window size used in the PIV computation. For each period $n$ the velocity field is computed from these two images, and the resulting set of velocity fields is averaged over 100 periods.} Since the measurements are performed in the laboratory frame, the measured velocity {field} includes the velocity of the cylinder, ${\bf u}_\mathrm{c}(t_n) = d {\bf r}_\mathrm{c} /dt (t_n)  = A \Omega [-\sin (\Omega t_n) {\bf e}_x + \cos (\Omega t_n) {\bf e}_y ] = - A \Omega {\bf e}_x$ (see Fig.~\ref{fig:wave_PIV}a). We simply deduce the wave velocity field in the cylinder frame by subtracting ${\bf u}_\mathrm{c}$ from the measured velocity fields.

Figure~\ref{fig:wave_PIV}(c,d) shows the wave fields for the two fluid viscosities $\nu = 50$ and 500~mm$^2$~s$^{-1}$, obtained for a forcing frequency $\Omega/\omega_{1} = 0.67$ and amplitude $A = 2.9$~mm ($\epsilon = 0.057$). The corresponding Reynolds numbers are $Re=660$ and 66, respectively. These wave fields are in good qualitative agreement with the potential solution, shown in Fig.~\ref{fig:wave_PIV}(b), except for two features: first, boundary layers are clearly visible at the cylinder wall, of typical thickness 4 and 12~mm respectively, which corresponds to $\simeq 2.5 \sqrt{\nu / \Omega}$. Second, we observe a significant phase delay between the wave velocity and the cylinder velocity (which is along $-{\bf e}_x$ at this phase), of order $\Delta \simeq 30^\mathrm{o}$ for the larger viscosity case. Such phase delay was also reported by Ducci {\it et al.}~\cite{Ducci2014}, for different fluid viscosity and aspect ratio.

\begin{figure}
    \centerline{
		\includegraphics[width=14cm]{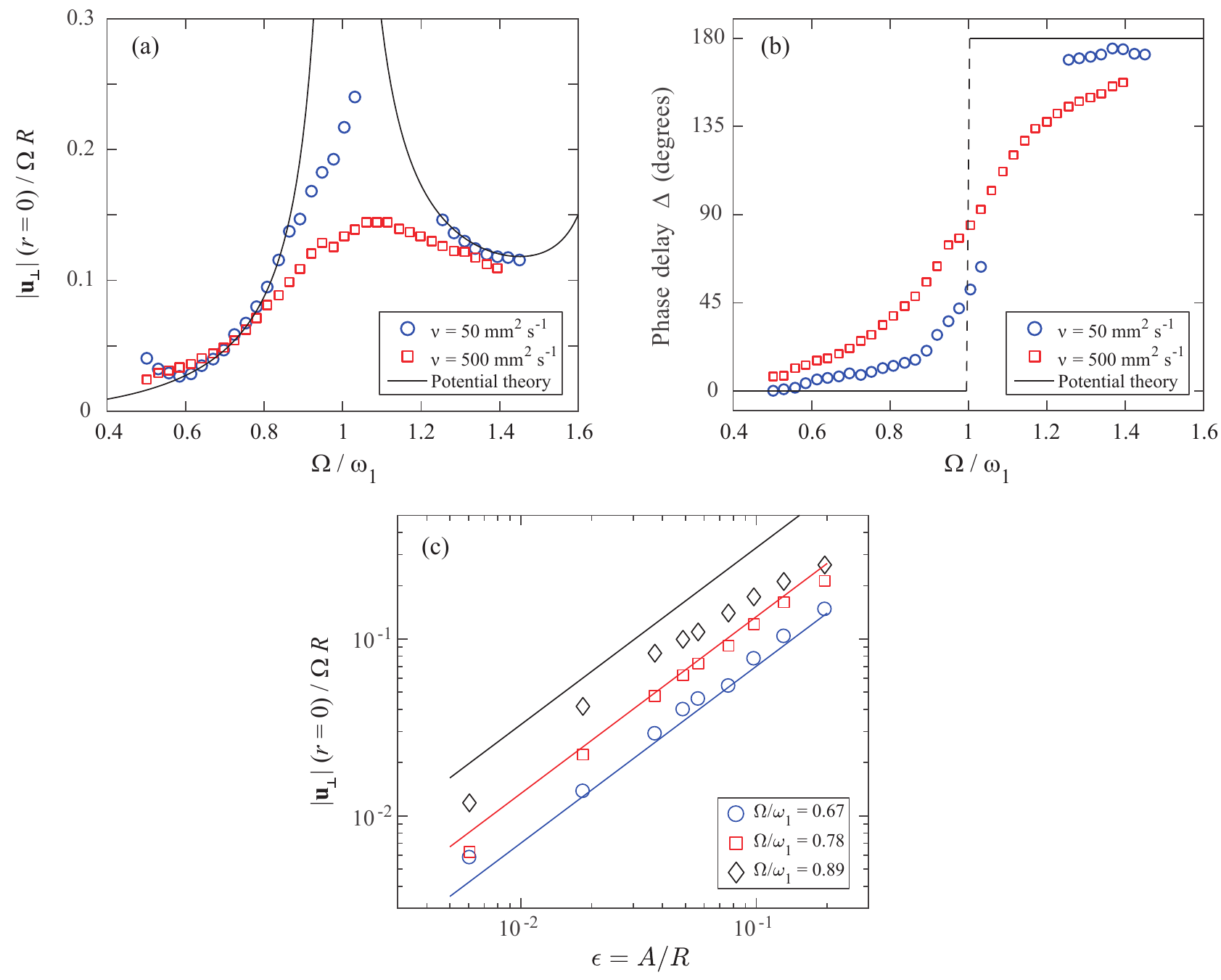}
		}
    \caption{(a) Norm of the wave velocity measured at $r=0$, $z_0/R = -0.23$, as a function of the forcing frequency $\Omega/\omega_1$ at a fixed forcing amplitude $\epsilon = 0.057$, for the two fluid viscosities. \cor{In the range $\Omega/\omega_1 \in [1.05, 1.25]$ for $\nu = 50$~mm$^2$s$^{-1}$ the wave amplitude is too strong for PIV measurement and is subject to hysteresis.}   (b) Phase delay $\Delta$ between the forcing and the wave field at $r=0$.   (c) Norm of the wave velocity as a function of the forcing amplitude $\epsilon$ at three values of the forcing frequency, for $\nu = 500$~mm$^2$s$^{-1}$. Solid lines show the potential theory prediction (\ref{eq:normupt}).}
		\label{fig:wavevelocity}
\end{figure}

The amplitude and the phase delay of the wave field have been systematically measured for a forcing frequency $\Omega / \omega_{1}$ in the range $0.5 - 1.5$, at a fixed forcing amplitude $\epsilon = 0.057$. Figure~\ref{fig:wavevelocity}(a) compares the wave amplitude, defined as the norm of the horizontal velocity at the center, $|{\bf u}_\perp|(r=0) = \sqrt{u_r^2+u_\theta^2}$,  to the potential theory,
\be
\frac{|{\bf u}_\perp| (r=0,z=z_0) }{\Omega R} = \epsilon \sum_{n=1}^{\infty} \alpha_n k_n \cosh [k_n(z_0+H)/R],
\label{eq:normupt}
\ee
obtained by taking $J_1(x) \simeq x/2$ for $x\rightarrow 0$ in Eq.~(\ref{eq:uth}).
A good agreement is found at moderate forcing frequency, in the narrow range $\Omega/\omega_1 \in [0.6, 0.8]$. For lower frequency, the determination of the wave velocity is limited by the subtraction of the cylinder velocity $|{\bf u}_c|/\Omega R = \epsilon$, which becomes larger than the wave velocity. At larger frequency, the measured wave amplitude is smaller than the potential prediction, which diverges at $\Omega/\omega_1=1$. For the larger viscosity, the divergence is clearly smoothed, with a maximum wave amplitude shifted at $\Omega/\omega_1 \simeq 1.1$. For the smaller viscosity, there is a range of frequency around the resonance, $\Omega/\omega_1 \in [1.05, 1.25]$, in which the wave amplitude is too strong and cannot be measured by PIV (the free surface intersects the laser sheet). Outside this range we observe an asymmetric resonance curve, which can be attributed to an hysteresis of the wave above the resonance (only measurements for increasing forcing frequency are shown), in good agreement with the results of Reclari {\it et al.}~\cite{Reclari2014}.

The phase delay $\Delta$ between the forcing and the wave velocity, defined as the angle between the fluid velocity at the center of the cylinder and $-{\bf e}_x$, is plotted in Fig.~\ref{fig:wavevelocity}(b) as a function of the forcing frequency. According to the potential theory, the wave is in phase with the forcing for $\Omega < \omega_1$ ($\Delta=0$), and out of phase for $\Omega > \omega_1$ ($\Delta=180^\mathrm{o}$). Here again, we observe a good agreement with the theory far from the resonance, and a viscous smoothing of the phase jump near $\Omega/\omega_1 \simeq 1$.

Finally, the linear scaling with respect to the forcing amplitude $\epsilon$ in Eq.~(\ref{eq:scaluw}) is checked by varying $\epsilon$ in the range $0.006-0.20$ at fixed forcing frequency $\Omega$. Results are plotted in Fig.~\ref{fig:wavevelocity}(c) for the most viscous fluid. We find a good agreement between experiments and the potential prediction (\ref{eq:normupt}) at moderate frequency, $\Omega/\omega_1= 0.67$ and 0.78, over the whole range of $\epsilon$. At  larger frequency, however, the wave amplitude is below the potential theory, and the linear increase with $\epsilon$ starts to saturate for the largest values of $\epsilon$.

\section{Mean mass transport}
\label{sec:mf}

\subsection{Interpreting stroboscopic PIV measurements}

\cor{We now turn to the  Lagrangian mean flow $\overline{{\bf u}}$ generated by the orbital sloshing wave in the range of parameters $(\epsilon, \Omega/\omega_1)$ for which the potential theory provides a reasonable description of the wave flow. This Lagrangian mean flow contains both the Eulerian steady streaming contribution $\overline{{\bf u}}_{str}$ and the Stokes drift contribution $\overline{{\bf u}}_{Sto}$, that cannot be discriminated by stroboscopic PIV measurements. It is instructive to see why.}

\begin{figure}
    \centerline{\includegraphics[width=8cm]{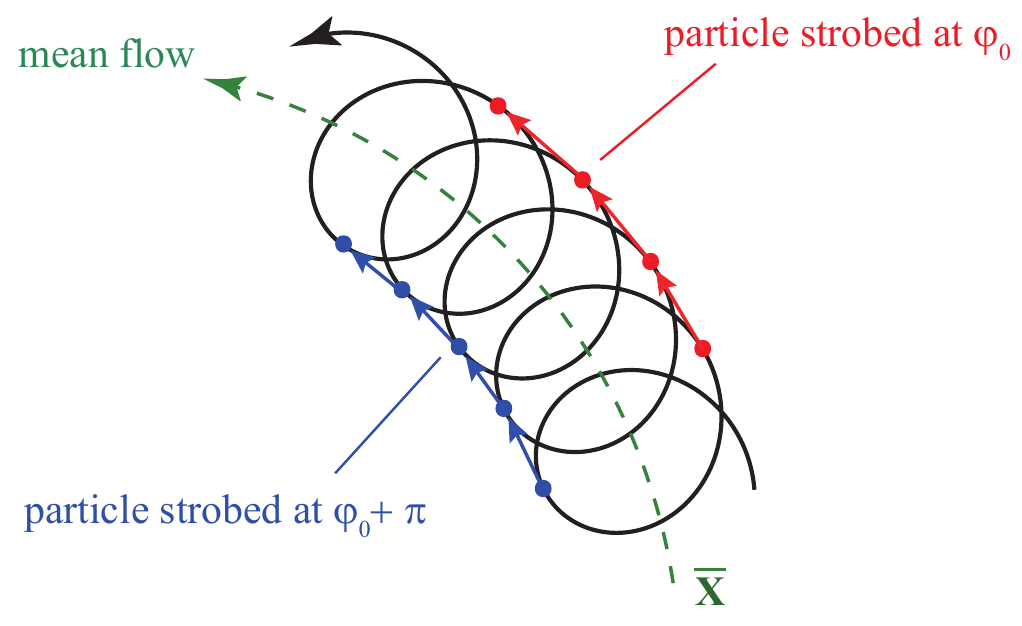}}
    \caption{In the weakly nonlinear regime, the particle trajectory (thick black line) is the sum of a strong oscillating flow ${\bf u}$, of order $\epsilon$, and a weak mean flow $\overline{\bf u}$ (dashed green line), of order $\epsilon^2$; the mean flow contribution is exaggerated here for clarity. When imaged at a given phase of the wave (here $\varphi_0$ in red and $\varphi_0  + \pi$ in blue), stroboscopic PIV introduces a systematic bias  of order $\epsilon^3$ in the determination of the mean flow. The true mean trajectory $\overline{\bf X}(t)$ can be reconstructed by averaging the particle displacements over the two phases.}
		\label{fig:spiv}
\end{figure}

Let   ${\bf u}  + \overline{{\bf u}}_{str} $ be the total Eulerian velocity field, composed of the time-periodic wave flow  ${\bf u} ({\bf x},t)$ of order $\epsilon$ and the steady streaming flow  $\overline{{\bf u}}_{str} ({\bf x})$ of order $\epsilon^2$.  We consider a particle that follows the path ${\bf X} (t)$ and is imaged at positions ${\bf X} (t_n), {\bf X} (t_{n+1}) $ at times $t_n, t_{n+1}$ separated by one period $T=2 \pi / \Omega$ (see Fig.~\ref{fig:spiv}). The mean \cor{Lagrangian velocity of the particle, as measured by stroboscopic PIV,} then corresponds to
\be \label{eq:PIVth}
{\bf u}_{\mathrm{SPIV}} = \frac{{\bf X}(t_{n+1})  - {\bf X}(t_{n}) }{T} = \frac{1}{T} \int_{t_n}^{t_{n+1}}   \left [ {\bf u} ({\bf X} (t), t )   + \overline{{\bf u}}_{str} ({\bf X} (t) )  \right ] \  d t.
\ee
We introduce a first order estimate of the trajectory
\be
{\bf X} (t)  \approx \overline{{\bf X}}   + {\bf \eta} (\overline{{\bf X}} , t  )  +   O (\epsilon^2)  \quad , \quad {\bf \eta} (\overline{{\bf X}} , t  )  = \int_{t_n}^t {\bf u}  (\overline{{\bf X}} ,t' ) d t'.
\ee
Here $\overline{{\bf X}}$ is the average particle position over that period and the field $ {\bf \eta}$ is the particle excursion of order $\epsilon$, around this mean position.  We use this estimate of ${\bf X} (t)$ to reexpress the integrand of \eqref{eq:PIVth} using a Taylor expansion around the mean position $\overline{{\bf X}}$. Using the periodicity of ${\bf u} (\overline{{\bf X}} , t )$ and  ${\bf \eta} (\overline{{\bf X}}, t  ) $  we find that 
\be \label{eq:SPIV}
{\bf u}_{\mathrm{SPIV}} =\overline{{\bf u}}_{str} (\overline{{\bf X}})   +  \underbrace{ \frac{1}{T} \int_{t_n}^{t_{n+1}} {\bf \eta} \cdot \nabla {\bf u} |_{\overline{{\bf X}}} \, dt } _{\overline{{\bf u}}_{Sto} (\overline{{\bf X}})  }  +   O (\epsilon^3).
\ee
The leading order contribution is therefore the sum of the streaming velocity $\overline{{\bf u}}_{str} $ and the Stokes drift contribution $\overline{{\bf u}}_{Sto} $, which cannot be separated in stroboscopic PIV measurements.

\subsection{Theoretical predictions for $\overline{{\bf u}}_{Sto}$ and $\overline{{\bf u}}_{str}$}

The Stokes drift $\overline{{\bf u}}_{Sto}$ far from the boundaries can be computed to a good approximation from the inviscid wave solution~\cite{Hutton1964}.  Denoting ${\bf u}=  {\bf v} \exp (i \Omega t ) + c.c. $, we have ${\bf \eta}=  {\bf v} \exp (i \Omega t ) / (i \Omega)  + c.c. $, and the Stokes drift can be simply expressed as
\be
\overline{{\bf u}}_{Sto} = \frac{2}{\Omega} \mbox{Im} ({\bf v} \cdot \nabla {\bf v}^* ),
\ee
where $\mbox{Im}$ stands for imaginary part and the stars indicate complex conjugate. Using this formula, it is easy to see that only the azimuthal component of the drift,
\be 
\overline{u}_{Sto,\theta}  =  \frac{2}{\Omega} \mbox{Im} \left ( v_r    \frac{\pd  v_\theta^* }{\pd r}  +  \frac{v_\theta }{r}  \frac{\pd v_\theta^* }{\pd \theta}   + v_z \frac{\pd v_\theta^*}{\pd z}  +  \frac{v_\theta v_r^*}{r} \right ) \nonumber \\
\ee
is non-zero and equal to
\ba \label{eq:sdrift}
\overline{u}_{Sto,\theta}
&=&  \epsilon^2 \left \{  \left [  \sum_{n=1}^{+\infty} \alpha_n \frac{k_n^2}{2} [ J_1 (...) + J_3  (...)  ]  \cosh(...) \right ]  \left [  \sum_{n=1}^{+\infty} \alpha_n k_n  J_2 (..)  \cosh(..) \right ] \right . \nonumber \\
&  &\hspace*{.3cm}  \left . +   \left [  \sum_{n=1}^{+\infty} \alpha_n k_n^2 [ J_0 (...) + J_2  (...)  ]  \sinh(...) \right ]  \left [  \sum_{n=1}^{+\infty} \alpha_n k_n  J_1 (...)  \sinh(...) \right ]
 \right \},
\ea
where the omitted arguments of the Bessel and \cor{hyperbolic} functions are $k_n r /R$ and $k_n (z+H)/R$, respectively. This  purely azimuthal inviscid Stokes drift indicates that the poloidal recirculations found in the  experiment can only be due to the steady streaming flow or to viscous corrections of the Stokes drift. 

\cor{The steady streaming flow $\overline{{\bf u}}_{str}$ is difficult to calculate and requires a dedicated analysis that is out of the scope of the present article. It is however useful to get an idea about the structure of this calculation. Streaming is the reaction of the flow to time-averaged non-linear stresses $\overline{{\bf u} \cdot \nabla {\bf u}}$, but since inviscid potential waves cannot induce mean vorticity ($\nabla \times ( {\bf u} \cdot \nabla {\bf u} ) = \nabla \times (\nabla | {\bf u} |^2 / 2 ) = {\bf 0}$ for inviscid potential waves) it is necessary to consider nonlinear interactions in the viscous boundary layers to find the origin of the streaming.}

\cor{In the orbital shaking problem, there are 4 different boundary regions in which the viscous boundary layers strongly differ: (a) near the rigid walls; (b) near the free surface; (c) in corners of the rigid walls; (d) near the contact line. In practice, only the layers in regions (a) and (b) are analytically tractable using multi-scale expansions, and regions (c) and (d) are almost never considered.  In the most advanced theoretical models on streaming induced by oscillatory boundary layers under waves (see e.g. \cite{Nicolas2003,Martin2006,Perinet2017}), non-linear interactions in regions (a) and (b) are calculated using matched asymptotics. This gives rise to a set of boundary conditions that serve as input to solve numerically the Craik-Leibovich equation \cite{Craik76} in the bulk of flow. In the weak streaming limit inertia can be neglected and this equation reduces to a simple Stokes problem. In the strong streaming limit, however, all nonlinear terms, even the nonlinear interaction that involves the Stokes drift, must be accounted for.}

\cor{A major difficulty in computing the streaming flow in the orbital sloshing configuration is the complexity of the flow in region (d) near the contact line, where the wave amplitude is the largest. The natural choice of a stress-free condition at the surface with no pinning of the contact line at the wall is extremly difficult to achieve experimentally: even slight surface contamination may dramatically change the boundary conditions and hence the resulting streaming flow~\cite{Martin2006,Perinet2017}. Although the use of silicon oil in the present experiment is expected to minimize surface contamination effects, a pure stress-free condition can certainly not be guaranteed.}

\subsection{Phase bias in stroboscopic PIV measurements}

\begin{figure}
    \centerline{\includegraphics[width=14cm]{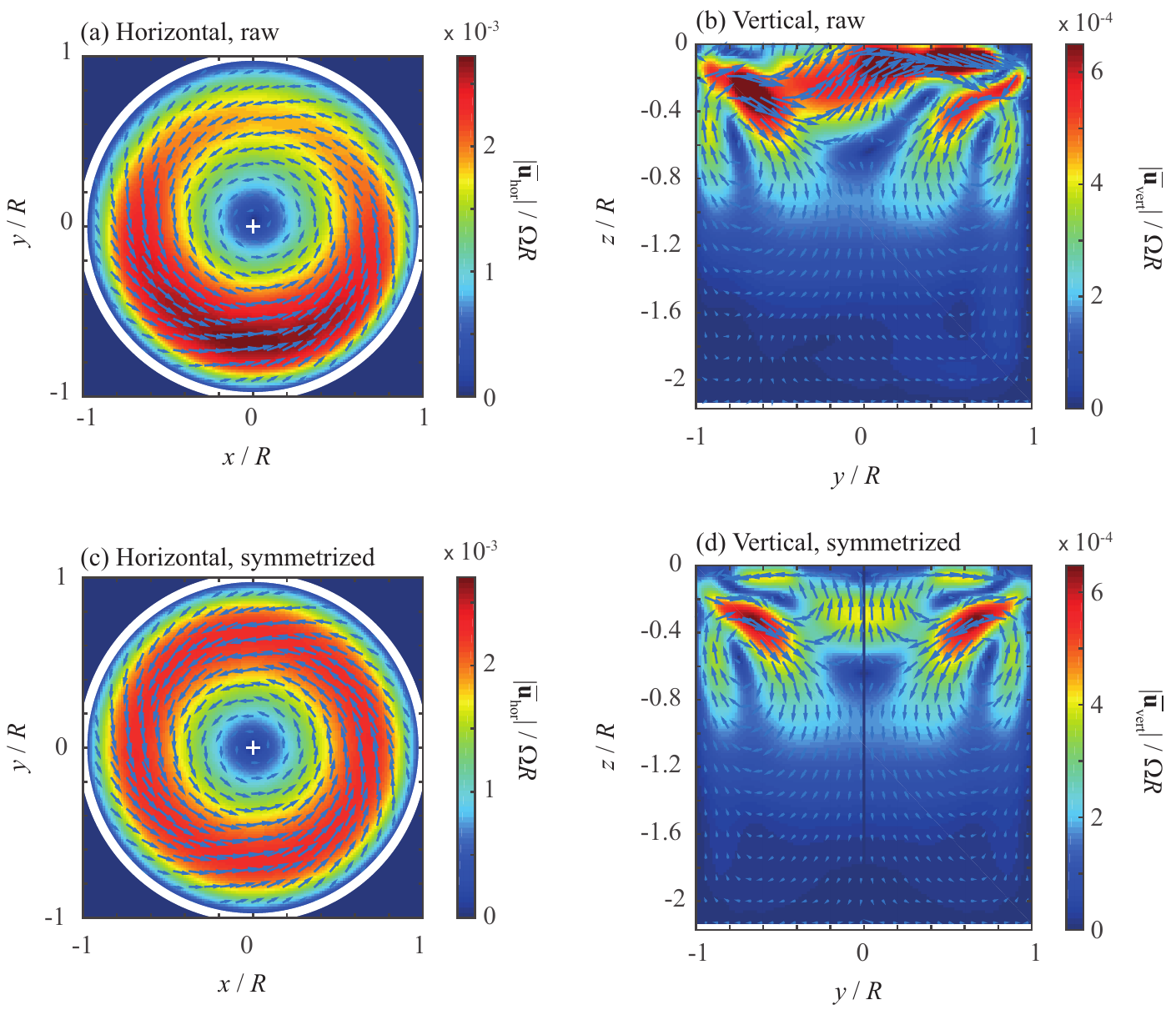}}
    \caption{Mean flow measured by stroboscopic PIV at phase $\varphi_0 = \pi/2$, (a,c) in the horizontal plane at $z_0/R = -0.23$ below the free surface, and (b,d) in the vertical plane. Forcing frequency $\Omega/\omega_1=0.67$, amplitude $\epsilon = 0.057$, viscosity $\nu = 500$~mm$^2$~s$^{-1}$. The upper line (a,b) shows the raw velocity fields, and the lower line (c,d) shows the symmetrized velocity fields using Eq.~(\ref{eq:symm}) to account for the wave contribution.}
		\label{fig:meanflow_sym}
\end{figure}

The mean flow measured by stroboscopic PIV is shown in figure~\ref{fig:meanflow_sym}, in the horizontal plane (at a distance $z_0/R = -0.23$ below the free surface) and in the vertical plane. The mean flow is mainly azimuthal, but with a slightly off-centered minimum velocity.  The recirculation in the vertical plane, mostly active in the upper half of the cylinder, is composed of a nearly axisymmetric toroidal vortex, ascending near the cylinder wall and descending along the axis, together with a strong non-axisymmetric surface current. For these parameters ($\Omega/\omega_1=0.67$, $\epsilon = 0.057$), the azimuthal velocity is $\simeq 2 \times 10^{-3} \, \Omega R$, and the radial and vertical velocities are $\simeq 6 \times 10^{-4} \, \Omega R$: as expected this mean flow is much smaller than the wave flow, which is of order of $5 \times 10^{-2} \, \Omega R$ here (see Fig.~\ref{fig:wavevelocity}).

The axisymmetry breaking in the mean flow originates from a bias in the stroboscopic PIV method, which we illustrate in Fig.~\ref{fig:spiv}: the particle displacement during one oscillation period contains, for a non-homogeneous wave flow, a contribution which depends on the arbitrary acquisition phase $\varphi_0$. This highlights a real subtlety in stroboscopic PIV: although the previous manipulations \eqref{eq:SPIV} show that ${\bf u}_{\mathrm{SPIV}}$ provides a measure of the Lagrangian mean flow, we do not precisely know where that mean flow vector should attach, in ${\bf \overline{X}}$, ${\bf X}_n$, ${\bf X}_{n+1}$ or somewhere in between. Although the mean particle position ${\bf \overline{X}}$ is the most defendable choice, image correlation naturally locates the vector ${\bf u}_{\mathrm{SPIV}} $ at the phase-dependent particle positions, say at ${\bf X}_n$. Constructing the PIV field from these pointwise measurements therefore introduces a systematic bias, which can be expressed mathematically through the following Taylor expansion
\ba
{\bf u}_{\mathrm{SPIV}} ({\bf X}_n  ) &=& {\bf u}_{\mathrm{SPIV}} (\overline{{\bf X}} + {\bf \eta} (\overline{{\bf X}} , t_n  )  + O (\epsilon^2)) \nonumber \\
& = &  {\bf u}_{\mathrm{SPIV}} (\overline{{\bf X}})  + {\bf \eta} (\overline{{\bf X}} , t_n  )  \cdot \nabla  {\bf u}_{\mathrm{SPIV}} |_{\overline{{\bf X}} } + O (\epsilon^2). \label{eq:bias}
\ea
Attaching the mean flow vector to the instantaneous position ${\bf X}_n$ rather then to the mean position $\overline{{\bf X}}$ pollutes the mean flow measurement with a systematic bias ${\bf \eta} \cdot \nabla  {\bf u}_{\mathrm{SPIV}}$ that relates to the particle displacement ${\bf \eta}( \overline{X},t_n)$ and to the spatial gradient of the mean flow field in that direction. With $ {\bf u}_{\mathrm{SPIV}}$ of order $\epsilon^2$, the bias ${\bf \eta}  \cdot \nabla  {\bf u}_{\mathrm{SPIV}}$  is of order $\epsilon^3$. In principle, it is possible to reconstruct the field $ {\bf u}_{\mathrm{SPIV}} (\overline{{\bf X}}) $ by averaging stroboscopic measurements at many different phases. This costly procedure is however technically impossible in our experiment because the PIV setup is in the laboratory frame and only the two phases when the cylinder axis crosses the laser sheet can be measured.

\begin{figure}
    \centerline{
		\includegraphics[width=8cm]{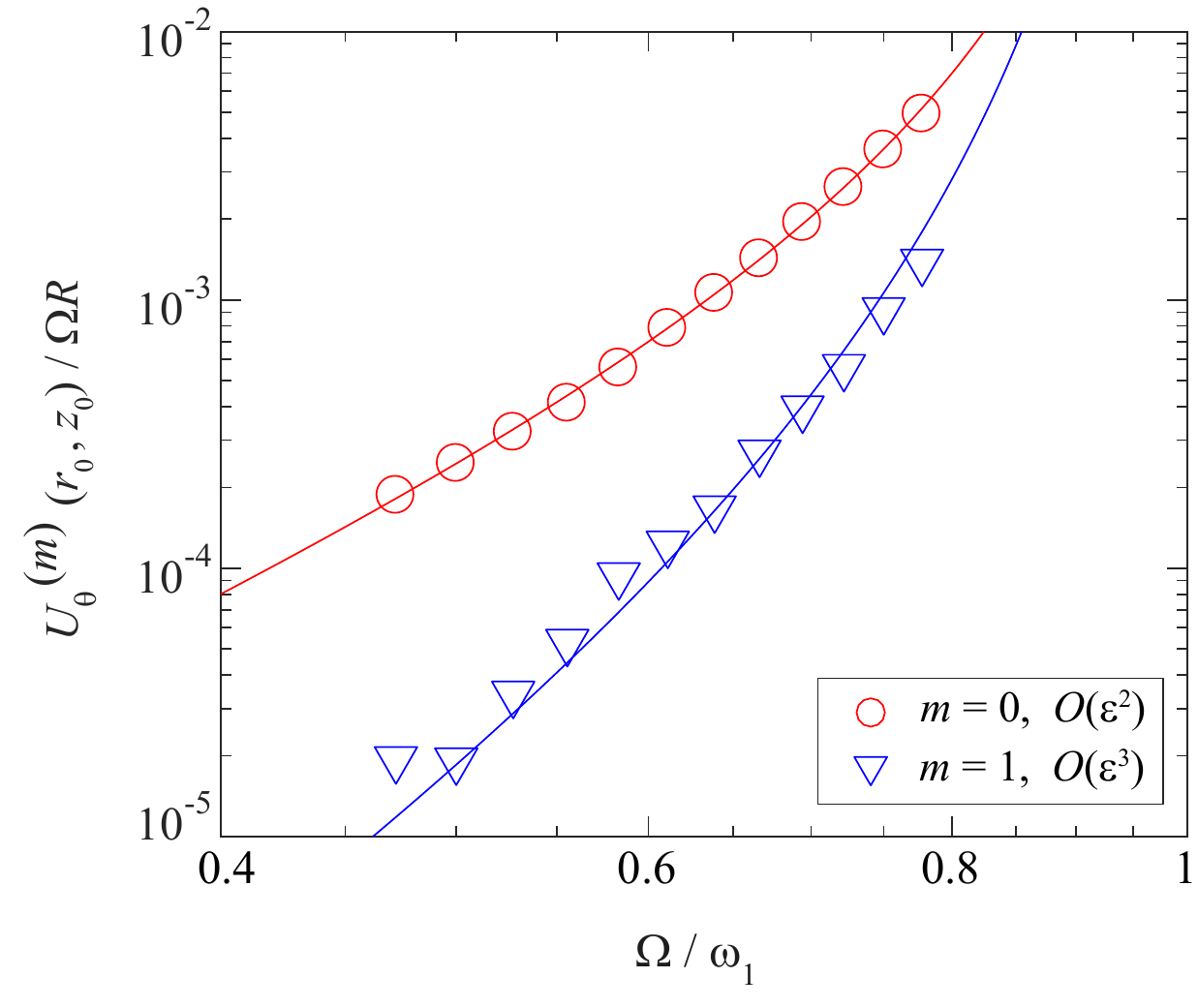}
	}
    \caption{Amplitude of the modes $m=0$ ($\circ$) and $m=1$ ($\triangle$) of the azimuthal mean flow at radius $r_0 = R/2$ and height $z_0 = -0.23 R$, as measured by stroboscopic PIV, as a function of the forcing frequency $\Omega/\omega_1$ at a fixed forcing amplitude $\epsilon = 0.057$, for viscosity $\nu=500$~mm$^2$~s$^{-1}$. The lines show the best fits with the laws $\epsilon^2/[(\omega_1/\Omega)^2-1]^2$ and $\epsilon^3/[(\omega_1/\Omega)^2-1]^3$, respectively.    
    }
		\label{fig:eps3}
\end{figure}

Even though we cannot filter the bias a whole, we can still filter out the largest contribution to it, \cor{which is dominated by the azimuthal wavenumber $m=1$.} Given one biased SPIV field $ {\bf u}_\mathrm{SPIV} (r,\theta,z)$ obtained at any arbitrary phase we can calculate 
\be
 {\bf \overline{u}} (r,\theta,z)  = \frac{1}{2} \left( {\bf u}_\mathrm{SPIV} (r,\theta,z) + {\bf u}_\mathrm{SPIV} (r,\theta + \pi,z) \right).
\label{eq:symm}
\ee
In the horizontal plane, we simply rotate the measured field over $\pi$, and in the vertical plane we apply a mirror symmetry with respect to the axis $r=0$. This procedure cancels all odd $m$- contributions to the flow field, but leaves all even $m$ contributions invariant, reducing the bias to order $\epsilon^4$. The validity of this procedure is illustrated in Fig.~\ref{fig:eps3}. We have decomposed the azimuthal component of the raw SPIV field, measured along the radius $r_0=R/2$ at height $z_0 = -0.23 R$, as
$$
u_{\rm{SPIV},\theta}(\theta; r_0, z_0) = U_{\theta}^{(0)} + U_{\theta}^{(1)} \cos(\theta - \theta_0)
$$
(with $\theta_0$ an arbitrary phase). The two modal contributions, \cor{$U_{\theta}^{(0)}$ and $U_{\theta}^{(1)}$}, plotted as a function of the forcing frequency $\Omega/\omega_1$ at fixed forcing amplitude $\epsilon$, indeed show the expected scaling, $\epsilon^2/[(\omega_1/\Omega)^2-1]^2$ and $\epsilon^3/[(\omega_1/\Omega)^2-1]^3$. \cor{Applying Eq.~(\ref{eq:symm}) therefore conserves the leading axisymmetric $U_{\theta}^{(0)}$ contribution and removes the $U_{\theta}^{(1)}$ correction.}

The reconstructed mean flow fields, symmetrized using Eq.~(\ref{eq:symm}), are shown in Fig.~\ref{fig:meanflow_sym}(c,d). The {poloidal} recirculation flow in the vertical plane now appears as two vortices, \cor{an upper one with ascending flow along the axis (previously hidden by the wave contribution), and a lower one with descending flow along the axis}, separated by a stagnation point at $z/R \simeq -0.65$. Note the intense oblique jets that sprout from the contact line region. This picture shows that the radial component in the horizontal plane strongly depends on the height of the measurement plane. For the plane chosen here, $z_0/R = -0.23$, we have $u_r \simeq 0$ near the center and $u_r < 0$ at the periphery, but other situations may be encountered for other values of $z$ (in particular $u_r>0$ at the free surface).

\subsection{Dependence of the mean flow pattern with governing parameters}
\label{sec:dmfp}

We have systematically measured the mean flow for the two fluid viscosities, for a forcing amplitude $\epsilon = 0.057$. Here measurements are performed over a restricted range of forcing frequencies, $\Omega/\omega_1 \in [0.5, 0.8]$, for which the wave flow is well described by the linear scaling (\ref{eq:scaluw}) (see Fig.~\ref{fig:wavevelocity}).  The mean flow for higher frequencies could not be measured using stroboscopic PIV, because of the particles swept out of the measurement plane after one oscillation period.

\begin{figure}
    \centerline{\includegraphics[width=14cm]{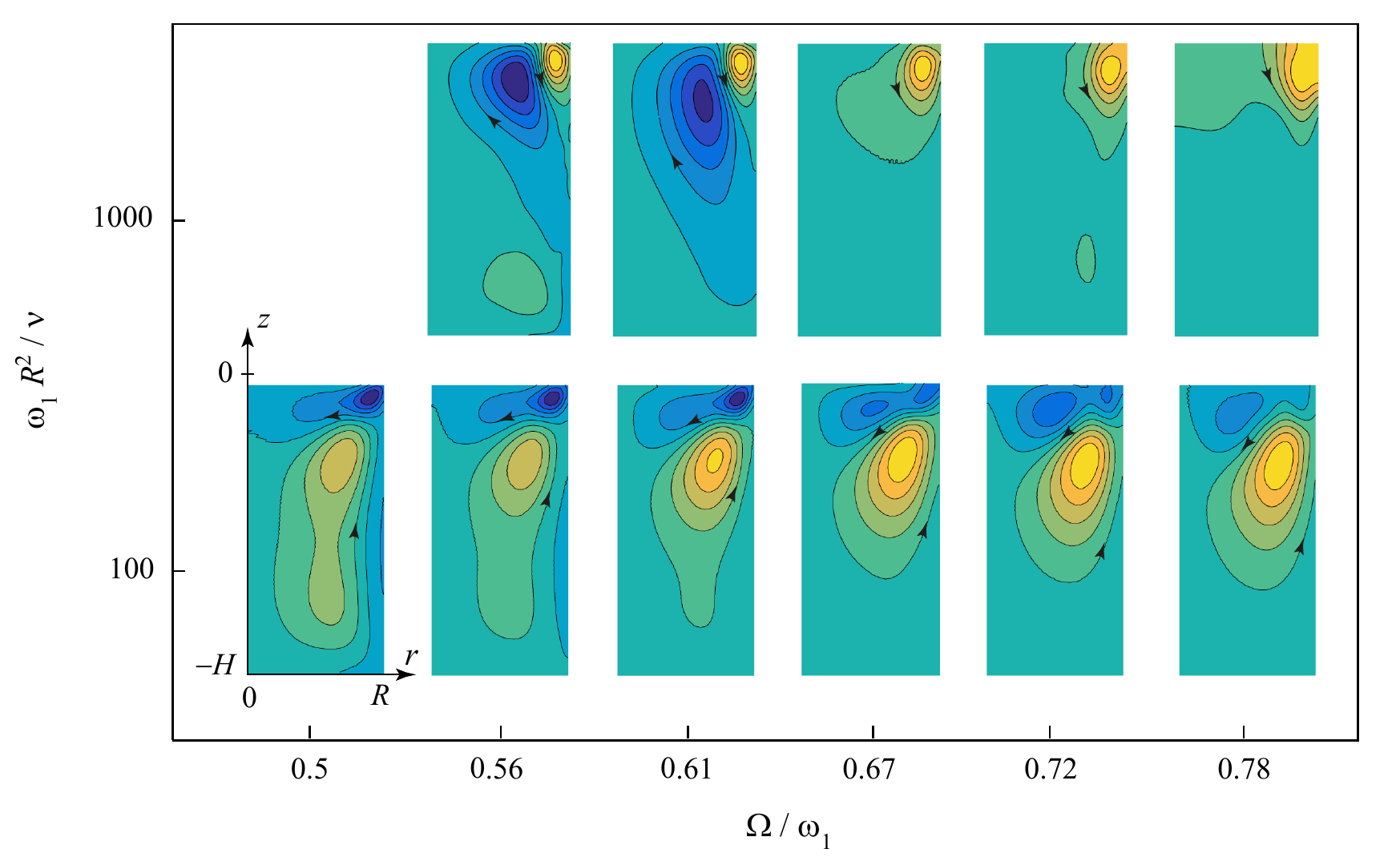}}
    \caption{Streamfunction of the mean flow in the vertical plane, for $\epsilon = 0.057$, for the two fluid viscosities, $\nu = 500$ and 50~mm$^2$~s$^{-1}$ ($\omega_1 R^2 / \nu = 100$ and 1000, respectively), showing the poloidal recirculation vortices active near the contact line.}
		\label{fig:streamfun}
\end{figure}

\begin{figure}
    \centerline{\includegraphics[width=14cm]{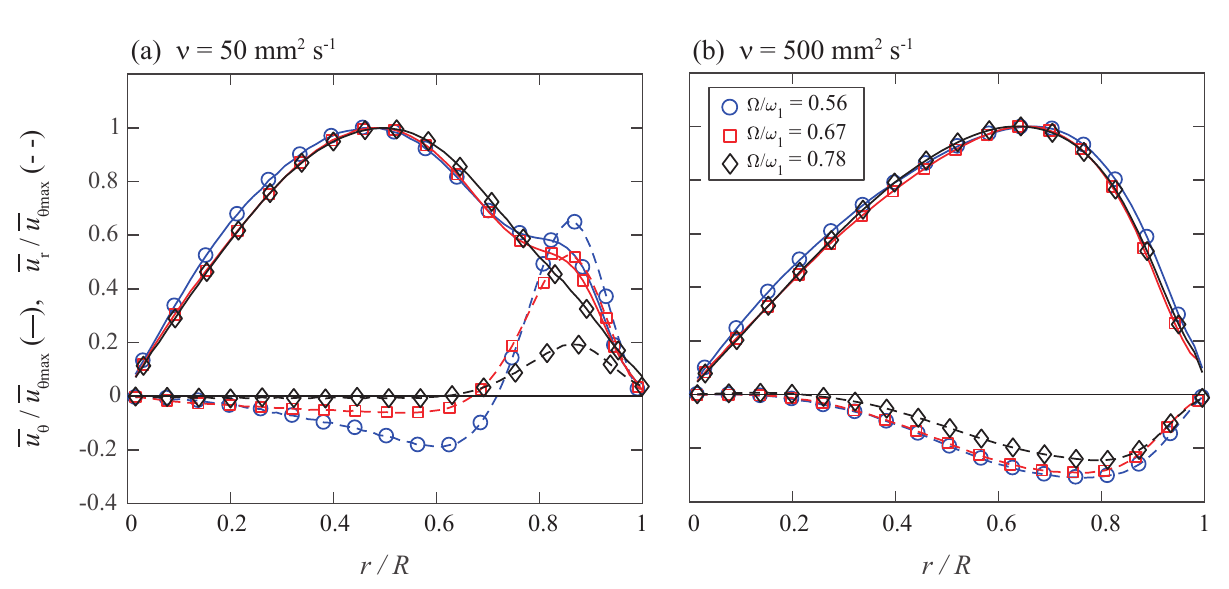}}
    \caption{Radial and azimuthal velocity profiles of the mean flow normalized by the maximum azimuthal velocity, measured at $z_0/R = -0.23$ for $\epsilon = 0.057$, (a) $\nu = 50$~mm$^2$~s$^{-1}$, (b) $\nu = 500$~mm$^2$~s$^{-1}$.}
		\label{fig:profils_umoy}
\end{figure}

The results are summarized in Fig.~\ref{fig:streamfun}, showing the axisymmetric (Stokes) streamfunction in the vertical plane, and in Fig.~\ref{fig:profils_umoy}, showing the radial profiles of the radial and azimuthal components normalized by the maximum azimuthal component at $z_0/R = -0.23$.  For both viscosities, the azimuthal velocity profiles are remarkably independent of the forcing frequency. The mean flow is nearly in solid-body rotation near the center, rotating in the direction of the orbital shaking: $\overline{\bf u} \simeq \overline{\omega}_0 r {\bf e}_\theta$ for $r/R<0.3$, with $\overline{\omega}_0>0$ the mean angular velocity. At larger radius, the mean azimuthal velocity decreases and the three velocity components become of the same order, marking the presence of a strong {poloidal recirculation} vortex near the contact line, where the wave amplitude is larger. This seems to be a very robust feature, at least for the fluid with lowest viscosity. A secondary {poloidal recirculation} vortex of weaker amplitude, located either below or at smaller radius than the primary vortex, is also present. The location and the rotation of these vortices are found to depend both on the viscosity and forcing frequency. As a result, the mean surface velocity near the wall may be either outward (mostly in the viscous case or at moderate forcing frequency) or inward, with the formation of a stagnation circle at the surface in some cases. 

Given that the structure of the wave flow does not change much in the studied range of parameters, it is unlikely that the weak streaming limit can explain the observed strong variation of the poloidal mean flow structure. Indeed, the (streaming) Reynolds number based on the mean flow, $Re_s = \overline{u} R/\nu $, is in the range $0.03-1$ for $\nu = 500$~mm$^2$s$^{-1}$, and $0.3-10$ for $\nu = 50$~mm$^2$s$^{-1}$, indicating that the criterion $Re_s \ll 1$ for the weak streaming limit is reasonably satisfied for the more viscous fluid, but not for the less viscous fluid. \cor{Accordingly, the variability of the complex patterns of poloidal streaming flow in Fig.~\ref{fig:streamfun} may originate from nonlinearities in the streaming flow.}

The strong dependence of the poloidal flow with the governing parameters makes quantitative comparison with the literature difficult.  We can note that the present results are in qualitative agreement with some previous experimental~\cite{Weheliye2013} and numerical~\cite{Kim2009} studies, although obtained for different aspect ratio and ranges of parameters. Weheliye {\it et al.}~\cite{Weheliye2013} observed, in a flat cylinder of aspect ratio $H/R = 0.6$ filled with water forced at $\Omega / \omega_1 = 0.56$ a single {poloidal} vortex with ascending velocity along the axis, compatible with the observed trend for increasing Reynolds number. In a cylinder of aspect ratio $H/R = 2$ and a weak forcing frequency of $\Omega/\omega_1 = 0.21$, Kim and Kizito~\cite{Kim2009} observed a similar toroidal vortex with ascending fluid along the axis at small viscosity ($\nu < 3$~mm$^2$~s$^{-1}$), and the formation of an additional counter-rotating {poloidal}  vortex near the wall at larger viscosity ($\nu > 10$~mm$^2$~s$^{-1}$), which again is compatible with our observations.

\begin{figure}
    \centerline{\includegraphics[width=14cm]{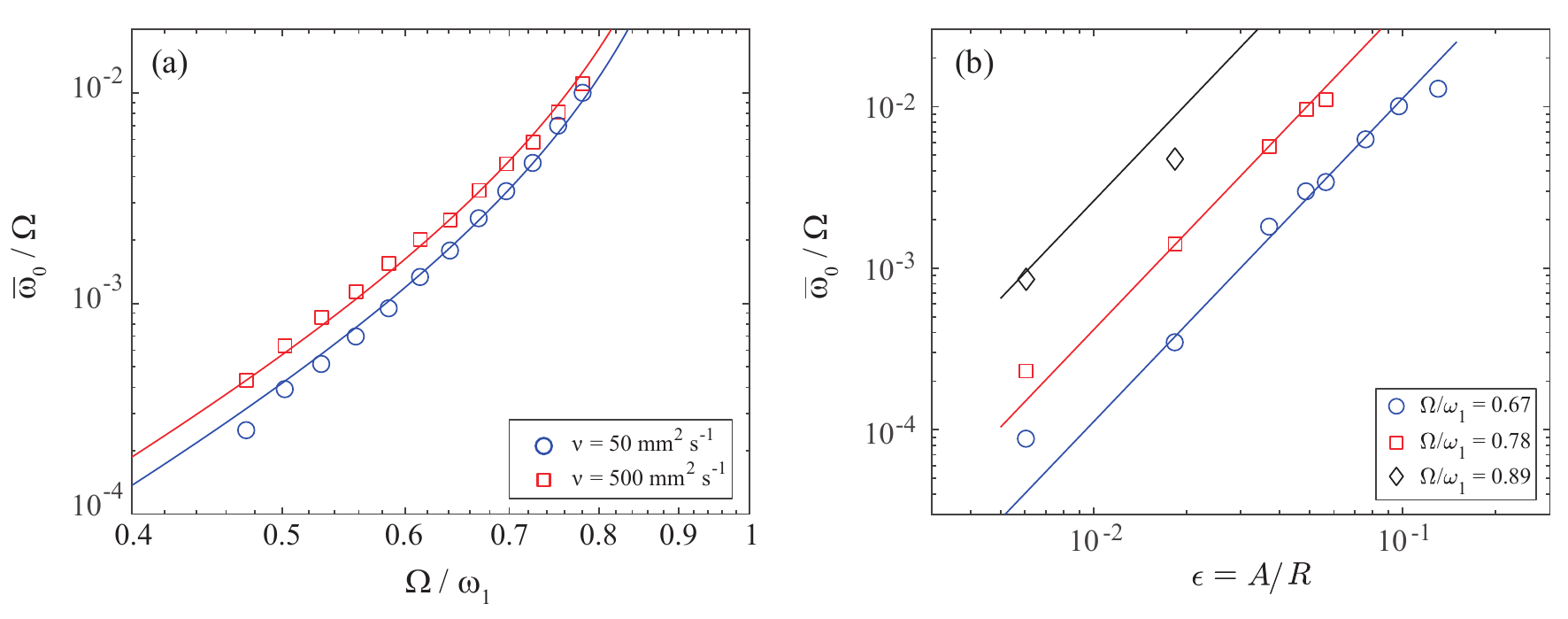}}
    \caption{Mean angular velocity $\overline{\omega}_0$ at $r=0$, $z_0/R=-0.23$, (a)  as a function of the forcing frequency $\Omega/\omega_1$ at a fixed forcing amplitude $\epsilon = 0.057$, for the two fluid viscosities; (b) as a function of the forcing amplitude $\epsilon$ at three values of the forcing frequency, for $\nu = 500$~mm$^2$s$^{-1}$. Solid lines show the weakly nonlinear scaling law (\ref{eq:scalw0}).}
		\label{fig:scalingmf}
\end{figure}

We now turn to the scaling of the mean flow {amplitude} as a function of the forcing frequency and amplitude. We characterize the mean flow by its dominant azimuthal contribution near the center: we compute the angular velocity, $\overline{\omega}_0 = \lim_{r \rightarrow 0} \overline{u}_\theta(r) / r$ at depth $z_0/R = -0.23$. The angular velocity, normalized by the forcing frequency $\Omega$, is plotted as a function of $\Omega/\omega_1$ in Fig.~\ref{fig:scalingmf}(a), and as a function of the forcing amplitude $\epsilon$ in Fig.~\ref{fig:scalingmf}(b). We first note that the range of variation of $\overline{\omega}_0$ is considerable: it varies over almost two orders of magnitude in the small range of forcing frequency explored here. Here again, measurement ranges are limited by the sweeping of the particles out of the measurement plane after one period for large values of $\epsilon$ and $\Omega$, and by residual thermal motion for small values of $\epsilon$ and $\Omega$. In practice, the largest measurable angular velocity using stroboscopic PIV is $\overline{\omega}_0 \simeq 10^{-2} \Omega$, i.e., the fluid performs one complete rotation after 100 forcing periods. 

The amplitude of the mean angular velocity is finally compared with the weakly nonlinear scaling law (\ref{eq:scalub}), written in the form
\be
\frac{\overline{\omega}_0}{\Omega} = K \frac{\epsilon^2}{[(\omega_{1}/\Omega)^2 - 1]^2},
\label{eq:scalw0}
\ee
with $K$ a tunable non-dimensional constant. In spite of its simplicity, this scaling provides a remarkable description of the data.  Viscosity is found to slightly enhance the mean flow, an effect which cannot be accounted for by Eq.~(\ref{eq:scalw0}): best fits with respect to $\Omega/\omega_1$ at fixed $\epsilon$, shown in Fig.~\ref{fig:scalingmf}(a), yield $K \simeq 1.1 \pm 0.1$ for $\nu = 50$~mm$^2$~s$^{-1}$ and $K \simeq 1.5 \pm 0.1$ for $\nu = 500$~mm$^2$~s$^{-1}$. The $\epsilon^2$ scaling is also tested in the case $\nu = 500$~mm$^2$~s$^{-1}$, in Fig.~\ref{fig:scalingmf}(b), using the same value of $K$ as in Fig.~\ref{fig:scalingmf}(a). A good overall agreement is obtained at moderate forcing frequency, whereas a mean flow slightly smaller than predicted is obtained as the resonance is approached, which is consistent with the weaker wave amplitude observed in Fig.~\ref{fig:wavevelocity}.

It is interesting to compare the fitted values of $K$ with the one we would get with the Stokes drift alone \cor{predicted for the potential wave solution} (i.e., with no Eulerian steady streaming). From Eq.~\eqref{eq:sdrift}, and retaining only the first term $n=1$, we have
\be
K_{Sto} \simeq \frac{k_1^4}{2 (k_1^2-1)^2 J_1^2(k_1)} \frac{\sinh^2 (k_1 (H+z_0)/R))}{\cosh^2 (k_1 H/R)} \simeq 1.25,
\ee
which is remarkably close to the experimental data, suggesting that the mean zonal circulation is predominantly due to the Stokes drift. This might also explain why this mean azimuthal flow is more robust than the the poloidal recirculation: The Stokes drift being a kinematic effect associated to the wave, it will be essentially not affected by the instabilities of the streaming flow.

\section{Conclusion}

In this paper we characterized the wave flow and the \cor{Lagrangian} mean flow in an orbitally shaken cylinder in the weakly nonlinear regime. The wave flow, measured by conventional PIV, shows a spatial structure in the bulk and a scaling in amplitude close to the potential prediction for forcing frequency far from the resonance, $\Omega/\omega_1 < 0.8$, except for a significant phase delay which increases with viscosity. The Lagrangian mean flow, measured by stroboscopic PIV, is composed of a robust global rotation near the center, and {poloidal recirculation} vortices mostly active near the contact line. Far from resonance, the amplitude of the central rotation is well described by a simple weakly nonlinear scaling law, quadratic in forcing amplitude, with a weak dependence on viscosity. \cor{This central rotation can be primarily attributed to the Stokes drift, whereas the poloidal recirculation flow is subject to a series of bifurcations}, with changes in the number of  vortices depending on the control parameters of the flow, suggesting unstable streaming flow.

Only scarce comparisons can be performed with literature at the moment, and a full description of the steady streaming flow, even in the weakly nonlinear regime, is not available yet. Such model would require to build the viscous sloshing modes of the container and to compute from them the momentum transfer from the boundary layer region to the bulk. This is a difficult task for three-dimensional flows with free boundaries~\cite{Nicolas2003,Perinet2017}, even in the weak streaming limit, $ \overline{u} R/\nu \ll 1$. Considering the complexity and variation of the steady streaming flows observed here, we anticipate that the weak streaming limit likely cannot capture all the observed patterns, and that a nonlinear Craik-Leibovich model~\cite{Craik76} is required.

\cor{Computing the streaming flow in the orbital sloshing flow is a formidable task, because it requires a proper description of the wave flow near the contact line. This specificity has been ignored in Refs.~\cite{Nicolas2003,Perinet2017}, but here our observations suggest that this contact line region, where the wave amplitude is the largest, might well be crucial in understanding the spatial structure of the poloidal recirculation. Another difficulty  may arise from possible surface contamination effects~\cite{Craik1982,Martin2006,Higuera2014} that can be difficult to control in experiments.  In Faraday wave experiments performed in water, Perinet {\it et al.}~\cite{Perinet2017} found strong differences between the observed mean flow patterns and the theoretical predictions, which they attributed to surface contamination. Although the use of silicon oil in the present experiments is expected to minimize surface contamination and pinning effects, it is known that pure free-slip boundary condition at the surface is extremely difficult to achieve, and that departure from this ideal situation may strongly influence the resulting steady streaming flow.}

\acknowledgments

We acknowledge A. Aubertin, L. Auffray, R. Kostenko and R. Pidoux
for their experimental help, and D. C\'ebron, M. Rabaud and A. Sauret for fruitful discussions.
F.M. acknowledges the Institut Universitaire de France.


\end{document}